\documentclass[twocolumn]{aa}
\usepackage{graphicx}
\usepackage{txfonts}
\input epsf
\begin{document}

\def\plotone#1{\centering \leavevmode \epsfxsize=\columnwidth \epsfbox{#1}}
\def\wisk#1{\ifmmode{#1}\else{$#1$}\fi}
\def\lt     {\wisk{<}}
\def\gt     {\wisk{>}}
\def\le     {\wisk{_<\atop^=}}
\def\ge     {\wisk{_>\atop^=}}
\def\lsim   {\wisk{_<\atop^{\sim}}}
\def\gsim   {\wisk{_>\atop^{\sim}}}
\def\kms    {\wisk{{\rm ~km~s^{-1}}}}
\def\Lsun   {\wisk{{\rm L_\odot}}}
\def\Zsun   {\wisk{{\rm Z_\odot}}}
\def\Msun   {\wisk{{\rm M_\odot}}}
\def\um     {$\mu$m}
\def\sig    {\wisk{\sigma}}
\def\etal   {{\sl et~al.\ }}
\def\eg	    {{\it e.g.\ }}
\def\ie     {{\it i.e.\ }}
\def\bsl    {\wisk{\backslash}}
\def\by     {\wisk{\times}}
\def\half   {\wisk{\frac{1}{2}}}
\def\third  {\wisk{\frac{1}{3}}}
\def\nwm2sr {\wisk{\rm nW/m^2/sr\ }}
\def\nw2m4sr{\wisk{\rm nW^2/m^4/sr\ }}
\def\mic    {\wisk{\mu{\rm m}}}
\def\deg    {\wisk{^{\circ}}}
   \title{Using peak distribution of the cosmic microwave background for WMAP and 
Planck data analysis: formalism and simulations}

   \author{C. Hern\'andez--Monteagudo\inst{1}
		\fnmsep\thanks{Formerly at: Facultad de Ciencias, 
              Universidad de Salamanca, Pza de la Merced,
										s/n, 37008, Spain},
          A. Kashlinsky\inst{2} and
		  F. Atrio--Barandela\inst{3}
          }


   \institute{\inst{1}Max-Planck Institute f\"ur Astrophysik, Postfach 1317. 
		D-85741, Garching bei M\"unchen, Germany\\
              \email{chm@MPA-Garching.MPG.DE}\\
       \inst{2}SSAI, Code 685, Goddard Space Flight Center, Greenbelt, USA 
									MD 20771\\
			  \email{kash@haiti.gsfc.nasa.gov}\\
   		\inst{3}Facultad de Ciencias, Universidad de Salamanca. 
		Pza de la Merced s/n, 37008, Spain\\
			  \email{atrio@usal.es}\\
  	}


   \abstract{We implement and further refine the recently proposed method (Kashlinsky, 
Hern\'andez-Monteagudo \& Atrio-Barandela, 2001 - KHA) for a 
time efficient extraction of the power
spectrum from future cosmic microwave background (CMB) maps. The method is  
based on the clustering properties of peaks and troughs of the Gaussian CMB
 sky.
The procedure takes only $\frac{1}{2}[f(\nu)]^2N^2$ steps where $f(\nu)$ is the
fraction of pixels with $|\delta T|\geq\nu$ standard deviations in the map of
$N$ pixels. We use the new statistic introduced in KHA, $\xi_\nu$, which 
characterizes spatial clustering of the CMB sky peaks of progressively 
increasing thresholds. The tiny fraction of the 
remaining pixels (peaks and troughs) contains the required information on the 
CMB power spectrum of the entire map. The threshold $\nu$ is the only parameter that determines the accuracy of the final spectrum. 
We performed detailed numerical simulations for parameters of the two-year WMAP and Planck CMB sky data
including cosmological signal, inhomogeneous noise 
and foreground residuals. In all cases we find that the method
can recover the power spectrum 
out to the Nyquist scale of the experiment channel.
We discuss how the error bars scale with $\nu$ allowing to decide between accuracy and speed.
The method can determine with significant accuracy the CMB power
spectrum from the upcoming CMB maps in only $\sim(10^{-5}-10^{-3})\times N^2$
operations. 
  

   }

   \maketitle
%

\section{\bf Introduction.}

The sub-degree structure of the CMB probes linear scales that were inside the
horizon at the last scattering epoch. The CMB fluctuations on these
scales carry a signature of causal processes during the last scattering and
thereby provide a very important constraint on the physics of the early Universe
and the models for structure formation. The most popular of these models is the
cold dark matter (CDM) set of models based on the inflationary paradigm for the
evolution of the early Universe. The models are very appealing, not only because
of their relative simplicity, but also because they provide a clear-cut
set of predictions that can be verified by observations. One (and perhaps the
most critical) of these predictions is the sub-degree structure of the CMB
anisotropies. In the framework of CDM models with adiabatic density fluctuations
the structure of the CMB power spectrum reflects the linear physics of sound
waves and initial density perturbations. While on scales outside the horizon at 
de-coupling  the CMB field preserves the initial
power spectrum, on sub-degree scales the interaction between the photon fluid
and matter leads to a series of acoustic peaks. The relative height, width
and spacing of these peaks depend on a final set of the cosmological parameters
($\Omega_{\rm total}, \Omega_\Lambda, \Omega_{\rm b}, h$) and also serve
to validate the cosmological CDM paradigm (see Hu \& Dodelson 2002 for a 
recent review). It is then important to measure the
sub-degree structure of the CMB with high accuracy.

After its first year of operation, the WMAP experiment has measured the Cosmic
Microwave Background (hereafter CMB) at five different frequencies 
with the maximal angular resolution of $\sim 0.21\degr$ and sensitivity close
 to $175~{\rm \mu K}$ per 7' pixel, (Bennet et al. 2003a). With these sensitivity and
angular resolution levels, the 
cosmological parameters as $\Omega_0$, $\Omega_{\Lambda}$, 
$\Omega_{\rm baryon}$, $H_0$, $n$, $\tau_{\rm reio}$ were determined with high accuracy (Spergel et al 2003). 
These measurements should be further improved with the planned ESA's mission
 Planck, where maps will contain up to $10^7$ pixels and extend to higher angular
resolution. Angular resolution and noise are both important for determining the 
cosmological parameters from the CMB.
   
A major challenge to understanding current and future CMB measurements is
to find an efficient algorithm that can reduce these enormous datasets:
$N\!\simeq\!10^5$ pixels in balloon experiments, $\simeq\!3\times 10^6$ for WMAP, 
to $\simeq\!5\!\times\!10^7$ for the Planck HFI data. (For comparison the COBE 
DMR data
analysis was based on only 4,144 pixels). Traditional methods require
inverting the covariance matrix and need $\sim\!N^3$ operations making them
unfeasible for the current generation of computers. 
Thus alternatives have been
developed for estimating the CMB multipoles from Gaussian sky maps:
Tegmark (1997) proposed a least variance
method that yields $C_l$'s directly from the temperature map in O($N^2$) operations,  
while Oh, Spergel \& Hinshaw (1999, hereafter OSH) perform an 
iterative maximization of the likelihood of the temperature map,
also in O($N^2$) operations. The first year WMAP results were analyzed with the OSH 
method (Hinshaw et al 2003). Bond, Jaffe \& Knox  (2000) and
Wandelt, Hivon \& G\'orski (2001) concentrate on the statistics of $C_l$'s, once 
the temperature map has been Fourier transformed. Hivon, et al. (2002) studied 
the likelihood of the power spectrum as obtained by direct Fast Fourier 
Transform of the available portion of the sky. Though it requires  
$O(N^{3/2}\ln N)$ operations, the accuracy of their results depends on the 
fraction and geometry of the sky covered.  A different approach consists in
computing the correlation function directly from the data in
 $O(N^2)$ operations. 
Smoot et al. (1992) used this type of analysis for
the first year release of the COBE data. More recently, Szapudi et al.
(2001), Szapudi, Prunet \& Colombi (2001)  have developed it for mega-pixel CMB data sets. The last
contribution in this field comes from Penn (2003), who applies
existing (Padmanabhan, Seljak \& Penn 2002) iterative algorithms on CMB data 
analysis,
and manages to estimate to power spectra in ${\cal O}$(N log N) operations.

A different method to compute the CMB power spectrum in a 
fast and accurate manner was proposed by us (Kashlinsky, Hern\'andez-Monteagudo \& 
Atrio-Barandela, 2001, hereafter KHA). The method exploits Gaussian 
properties of the CMB 
and noise fields and uses high peaks (and troughs) of the CMB field whose 
abundance is much smaller than the total number of pixels, $N$, and  whose 
correlation properties are strongly amplified in a way that depends on the 
underlying power spectrum.  This simultaneously achieves two important 
goals: reducing the number of computational steps for analysis and the 
good accuracy of the measured parameters. 
KHA have shown that the tiny fraction 
of the remaining pixels (peaks and troughs) contains the required 
information on the CMB power spectrum in the small scales. 
The peaks also trace, by default, the pixels with high signal-to-noise ratio 
and keep most of the information about the power spectrum of the signal.
Although this method may not provide as accurate results as other
more time consumings methods, it provides a new, non-standard and independent 
tool to unveil the sub-degree structure of the CMB.

In this paper we present detailed numerical simulations with the application of 
the KHA method to WMAP and Planck datasets in the presence of realistic 
components, such as inhomogeneous noise and non-Gaussian features expected 
from the Galactic foreground residuals. 
We show that with this method we can determine the CMB 
power spectrum outside the beam 
($l\!\simeq\! 640$ for WMAP and higher for Planck) in only $<\!10^{-3}N^2$ 
operations. We show that for the projected two-year WMAP noise levels our 
error bars at each $l$ are, 
on average, comparable to OSH, but have larger correlations which may 
be a small price to pay for the significantly shorter computational time.
For both WMAP and Planck 
parameters the KHA method is also faster than the direct computation of the 
CMB power spectrum in $O(N^{3/2}\log N)\sim N^{1.61}$ steps.

The structure of this paper is as follows: In Sect. 2 we briefly review the KHA formalism 
and in Sect. 3 we discuss its numerical implementation. Sect. 4 
deals with application of the method to both idealized and realistic WMAP
data. We show there that the method is immune to noise inhomogeneities 
and non-Gaussian features of Galactic foregrounds. Sect. 5 follows with results 
for application of the KHA method to Planck and in Sect. 6 we present our main conclusions. 

\section{\bf Mathematical formalism: an overview}

For completeness we give a brief overview of the KHA method; more details are
 in KHA. 

The CMB sky is expected to be highly Gaussian (Komatsu et al 2003) and this property, Eq. 
(\ref{P_definition}) below, is also widely used in standard maximum likelihood 
methods. For a Gaussian ensemble of $N$ 
data points (e.g. pixels) describing the CMB data
$\delta \equiv T - \langle T \rangle$ one will find a fraction
$f(\nu)=$erfc($\nu/\sqrt{2})$ with $|\delta| \geq
\nu\sigma$, where $\sigma^2 = \langle \delta^2\rangle$ is the variance of 
the field
and erfc is the complementary error function. The fraction of peaks, $f(\nu)$, 
is a rapidly decreasing function for $\nu\gsim 1$ and is e.g. 
$f(\nu)=(4.5,1,0.1)\times
10^{-2}$ for $\nu=(2,2.5,3)$ respectively.
The joint probability density of finding two pixels
within $d\delta_{1,2}$ of $\delta_{1,2}$ and separated by the angular distance
$\theta$ is given by the bivariate Gaussian:
\begin{equation}
p(\delta_1,\delta_2) = \frac{1}{2\pi \sqrt{||\mbox{\boldmath$C$}||} }
\exp(-\frac{1}{2} \mbox{\boldmath$\delta$} \cdot  \mbox{\boldmath$C$}^{-1}
\cdot  \mbox{\boldmath$\delta$})
\label{P_definition}
\end{equation}
where $C$ is the covariance matrix of the temperature field. 
We model the temperature correlation function (or {\it
matrix}, if defined on a set of pixels) as
\begin{equation}
C(\theta_{ij}) = C^{\rm CMB} (\theta_{ij}) + 
			\langle {\cal N}_i {\cal N}_j\rangle 
\label{covariance}
\end{equation}
where $C^{\rm CMB}(\theta_{ij})$ 
and $\langle {\cal N}_i {\cal N}_j \rangle$ are the CMB and the noise 
correlation matrices, respectively. We further define the total variance of 
the map as $C_0 \equiv C^{\rm CMB}(0)+\langle {\cal N}^2\rangle $. 
The case of inhomogeneous noise, i.e., 
noise whose variance varies across the sky, will be discussed in Sect. 4.2.
Note that the power spectrum of the map is nothing but the Legendre transform
of $C(\theta)$, (see Eq.~(\ref{eq:cl}) below).

The distribution of peaks of a Gaussian field
is strongly clustered (Rice 1954, Kaiser 1984, Jensen \& Szalay 1986, 
Bardeen et al. 1986, Kashlinsky 1987). 
Their angular clustering can be 
characterized by 
the 2-point correlation function, $\xi$, describing the excess probability of 
finding 
two events at the given separation.
 The correlation function of such regions 
is:
\begin{equation}
\xi_\nu(\theta) = \frac{ 2 \int_{\nu\sigma}^\infty\int_{\nu\sigma}^\infty
[p(\delta_1,\delta_2)+p(-\delta_1,\delta_2)] d\delta_1d\delta_2}
{[2\int_{\nu\sigma}^\infty p(\delta) d\delta]^2} - 1 = A_\nu(\frac{C}{C_0})
\label{xi}
\end{equation}
where $A_\nu$ is evaluated in detail in KHA to be:
\begin{equation}
A_\nu (x)=\frac{1}{H^2_{-1}(\frac{\nu}{\sqrt{2}})}
\sum_{k=1}^\infty \frac{ x^{2k} }{ 2^{2k}(2k)!
}H_{2k-1}^2(\frac{\nu}{\sqrt{2}})
\label{amplification}
\end{equation}
Here $H_n(x)=(-)^n \exp(x^2)d^n/dx^n \exp(-x^2)$ 
is the Hermite polynomial; $H_{-1}(x)\equiv \frac{ 
\sqrt{\pi} }{2}
\exp(x^2)$erfc($x)$.

At each angular scale the value of $\xi_\nu$ for every $\nu$ is 
determined
uniquely by $C$ at the same angular scale. In the limit of the entire
map ($\nu$=0) our statistic is $\xi_\nu$=0 and our method becomes meaningless; 
the new statistic has meaning only for sufficiently high $\nu$. One should 
distinguish between the 2-point correlation function, $\xi$, we directly 
determine from the maps, and the commonly used statistics in CMB studies, the 
temperature correlation function, $C$.

KHA thus suggested the following procedure to determine the power
spectrum of CMB in only $\simeq\! f^2(\nu)N^2\! \ll\! N^2$ operations:

1. Determine the variance of the CMB
temperature, $C_0$, from the data in $N$ operations;

2. Choose sufficiently high value of $\nu$ when $f(\nu)$ is small but
at the same time enough pixels are left in the map for robust measurement of
$\xi_\nu(\theta)$;

3. Determine $\xi_\nu(\theta)$ in $[f(\nu)]^2 N^2$ operations.

4. Finally, 
solve the equation
$A_\nu(C/C_0)\!=\!\xi_\nu(\theta)$ to obtain $C^{\rm CMB}(\theta)$ and from 
it $C_l$. This last step assumes that the experimental noise correlation
function $\langle {\cal N}_i {\cal N}_j \rangle$ has been properly determined, 
at least in the
set of pixels selected in the analysis, so that it can be subtracted from
the total correlation function $C(\theta_{ij})$. 

Formally speaking this procedure would require $\frac{1}{2} [f(\nu)]^2 N^2$
operations because of the symmetry in counting pairs. For simplicity we will be
referring to this as O$(f^2N^2)$ method implying a gain factor of $[f(\nu)]^2$
compared to other O$(N^2)$ methods.

\section{\bf Numerical implementation}

\subsection{Modeling the CMB sky}

Maps of the CMB sky were generated using the
hierarchical equal area isolatitude pixelization of the celestial sphere
implemented in HEALPix\footnote{HEALPix's URL site: 
{\it (http://www.eso.org/science/healpix/)}}.
Pixels have equal area and are arranged in ``constant latitude rings''.
Maps can be constructed with varying resolution, the number
of pixels given by $N = 12 \times N_{\rm side}^2$, being $N_{\rm side}$
the number of times in which each side of a pixel will
be divided in two, starting from a given initial configuration.
For a value of $N_{\rm side}$ of 512, one generates a map of 3,145,728 pixels
 of size of seven arcminutes. In the case of PLANCK, we used 
$N_{\rm side}=1024$ or 12,582,912 pixels of 3.5 arcminutes size.
The CMB spectrum was obtained using the CMBFAST code 
(Seljak \& Zaldarriaga, 1996).  

 For WMAP and PLANCK, we performed two types of simulations including: (1) cosmological 
signal plus white noise, and (2) cosmological signal, foregrounds and
inhomogeneous white noise. We chose three different thresholds:
$\nu=2.0, 2.27, 2.525$ for WMAP and $\nu=2.55, 2.78, 3.0$ for Planck.
At each threshold, the pixels selected are (on average) $f(\nu)N$
and their number doubles with decreasing threshold. The thresholds
for WMAP and Planck were chosen such that they select the same 
number of pixels in both experiments. In this way, we can analyze
the effect of pixel number on the accuracy of the method, i.e.,
we can estimate how the multipole error bars scale with threshold.

 From the peak spatial distribution, $\xi_{\nu}(\theta )$ was
computed on a grid of 31,415 equally spaced bins.
The power spectrum is computed using a Gauss-Legendre integration:
\begin{equation}
C_l = 2\pi \int_0^{\pi} d\theta \sin \theta C(\theta ) P_l (\cos \theta ).
\label{eq:cl}
\end{equation}
Accurate integration requires the evaluation of $C(\theta)$ 
at the roots of the Legendre polynomial out to a maximum order $l_{\rm max}$ 
(Press et al 1992) and this property was used by Szapudi et al (2001),
Szapudi, Prunet, \& Colombi (2001) and KHA. The 
value of $l_{\rm max}=800$ was chosen as a compromise between
the beam scale and the pixel angular scale, in order 
to prevent the first root of the Legendre
polynomial from being within the pixel size where there is no information
on $\xi_{\nu}(\theta)$. Prior to inverting Eq.~(\ref{xi}) 
$\xi_{\nu}(\theta)$ was smoothed
using a Gaussian of width 4' centered on the roots of the Legendre polynomials.
We verified that other average schemes lead to essentially identical results.

\begin{figure}[h]
\begin{center}
\plotone{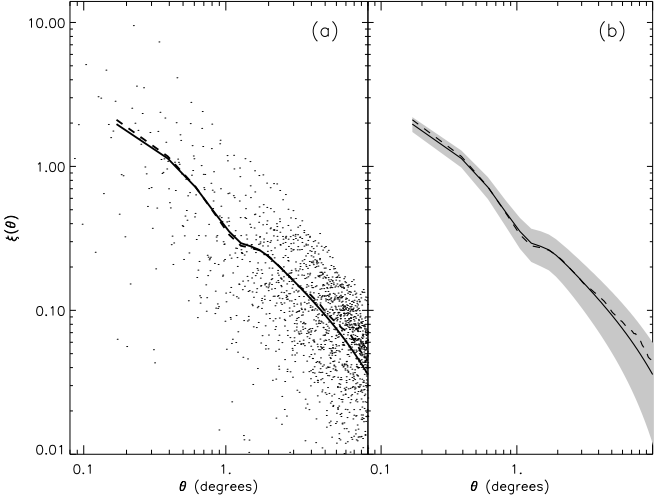}
\caption[]{(a) $\xi_{\nu=2}$ for the SCDM model with a Harrison-Zel'dovich 
power spectrum, obtained from a single realization of WMAP 94GHz channel.
Solid line: theoretical estimate (Eq.~[\ref{xi}]). Dots: raw data 
of $\xi_{\nu=2}$ evaluated at 31,415 angular bins distributed uniformly 
from $\theta=0\deg$ to $180\deg$. Dashed line: result of filtering the
raw data with a Gaussian of $4'$ width at the roots 
of the Legendre polynomial of order $800$.
(b) The shaded area represents the $1\sigma$
optimal variance error bar (eq~[\ref{ov}]) for the 94 GHz WMAP channel.
Dashed and solid lines have the same meaning as in (a).
}
\label{figure1}
\end{center}
\end{figure}

Fig.~(\ref{figure1}a) shows $\xi_{\nu=2}(\theta)$ for a SCDM model,
with  $\Omega_{\rm total} = 1$,  $\Omega_\Lambda = 0$, $\Omega_{\rm b}=0.04$,
$H_0=55\;{\rm km\;s^{-1}\;Mpc}$ and
scalar perturbation spectral index $n_S=1$. The solid line corresponds to 
the theoretical prediction and the dashed line is a smooth average obtained
from the data. The agreement is very good out to $\theta \sim 10\deg$, 
allowing an accurate  reconstruction of the power spectrum 
in the range of interest: $l\gsim 30$. 

\subsection{Accuracy of $\xi$ and correlations}

There are numerous ways to estimate $\xi_\nu(\theta)$ from the data, each way 
coming with its own uncertainty and systematics. Ideally, the uncertainty in 
$\xi_\nu(\theta)$ should reach $1/\sqrt{N_{\rm pairs}}$ 
but in practice systematic and other effects become 
dominant (Peebles 1980). Because the KHA method involves 
a complex and non-linear algorithm for inverting $C_l$'s from $\xi_\nu(\theta)$, 
the accuracy of the resultant CMB power spectrum may depend sensitively on 
the details, uncertainties and correlations of a particular estimator of $\xi_\nu$.

The ultimate accuracy with which the CMB power spectrum can be determined at
each $l$ is given by the so called optimal variance containing both the
cosmic variance and the instrument noise (Knox 1995):
\begin{equation}
\sigma_{\rm OV}(l) = \sqrt{\frac{2}{(2l+1)f_{\rm sky}}}
        [ C_l + \frac{4\pi  \langle {\cal N}^2 \rangle  B_l^{-2}}{N}].
\label{ov}
\end{equation}
In this equation, $N$ is the number of pixels, $f_{\rm sky}$ is the fraction
of the sky covered by the experiment and $B_l$ is the window function due to
the finite beam resolution.
This uncertainty also limits the accuracy with which one can 
determine the correlation functions $C$ and $\xi$. The latter are given by:
\begin{equation}
\sigma^2_{C(\theta )} = \sum_l | \frac{\partial C(\theta )}{\partial C_l} |^2
                \sigma^2_{\rm OV}(l), \quad
\sigma^2_{\xi_{\nu}(\theta )} =
        | \frac{\partial \xi_{\nu}(\theta)}{\partial C(\theta) }|^2
                \sigma^2_{C(\theta)}.
\label{eq:dxi}
\end{equation}
The shaded region in Fig.~(\ref{figure1}) shows an example of the optimal 
variance uncertainty in $\xi_{\nu=2}(\theta)$. (Note that it is a function of $\nu$).

In KHA, we evaluated $\xi_\nu$ as the ratio of number of peak pairs at a given angular
scale with respect to a poissonian catalog:
\begin{equation}
\hat{\xi_{\nu}}(\theta )=DD/RR - 1 .
\label{eq:xi}
\end{equation}
At each angular separation, $DD$ is the number of pairs of peaks and
$RR$ the number of pairs of $f(\nu)N$ points randomly located on the sphere
Landy \& Szalay (1993) argue that this estimator is neither optimal
nor unbiased. They proposed a more accurate estimator defined by:
\begin{equation}
\tilde{\xi_{\nu}}(\theta) =  (DD+RR-2DR)/RR ,
\label{eq:xinew}
\end{equation}
with $DR$ is the number of pairs given a cross-correlation between the 
$f(\nu)N$ peaks of the CMB map and the same number of
points Poisson distributed on the sky. 
We performed simulations for WMAP as described in
Sect. 4 and computed $\xi_\nu$ using both estimators. 
We found that, for the range of interest  ($\theta < 10\deg$), both
estimators show scatter equally close to Eq.~(\ref{eq:dxi}). Hence,
in what follows we shall use Eq.~(\ref{eq:xi}) to be consistent with KHA.

In Fig.~(\ref{figure1}b) we plot the 1-$\sigma$ errors in $\xi_{\nu=2}$ 
for the 94 GHz WMAP channel given by the noise and cosmic variances.
The smooth average obtained from the data (dashed line)
is in good agreement with theoretical value.
Therefore, we will expect that the radiation power spectrum 
obtained from $\xi_{\nu=2}$ as described below will be very close to 
that of the input model.  

\begin{figure}[h]
\begin{center}
        \plotone{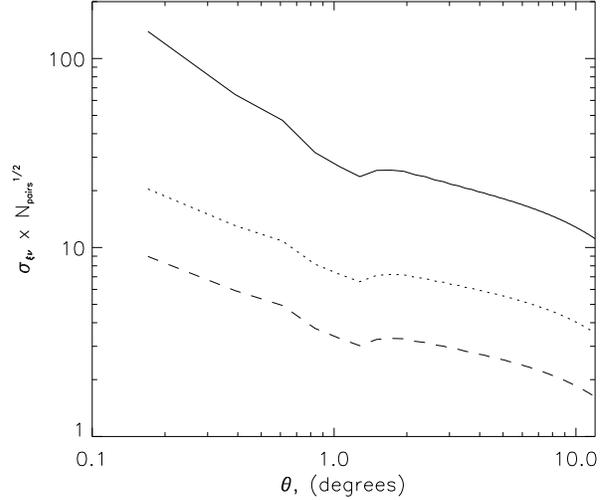}
\caption[]{The ratio of the optimal variance uncertainty on $\xi_\nu$ to 
the statistical uncertainty, $1/\sqrt{N_{\rm pairs}}$ plotted vs the 
separation angle for the resolution of the WMAP 94GHz channel. 
The solid line corresponds to 
$\nu=3$, dotted to $\nu=2$ and dashes to $\nu=1.5$.}
\label{figure2}
\end{center}
\end{figure}

The sampling variance associated with the low number of pairs - compared with 
methods based on the correlation function - is much smaller that the cosmic variance
on $\xi_\nu$.  Fig.~(\ref{figure2}) shows $\sqrt{N_{\rm pairs}}\sigma_{\xi,
{\rm OV}}$ which is 
the ratio of the optimal variance uncertainty on $\xi$ to the statistical 
uncertainty, $1/\sqrt{N_{\rm pairs}}$. The lines correspond to various 
$\nu$=1.5 (dashes), 2 (dots) and 3 (solid line). The ratio was computed for 
WMAP number of pixels. The plot shows that 
for any mega-pixel CMB map the method can determine the new statistic, 
$\xi_\nu$, out to the angles of interest optimally even for threshold as 
low as $\nu=1.5$. This is the reason why our results are not sensitive 
to the particular estimator used in calculating $\xi_\nu$.

At $\theta \gsim 10\deg-20\deg$ the value of $\xi_\nu(\theta)$ 
is small and its value becomes dominated by shot noise. Hence, 
we restrict the analysis to $\theta < 10\deg-20\deg$ by introducing a
taper function that cuts out the contribution of the correlation
function for $\theta \gsim 10\deg$. As a result (a) we will not 
recover very accurately multipoles below $l\simeq 30$ and (b) tapering will 
introduce additional correlations among the different $C_l$'s. The first limitation
is irrelevant, since one can always degrade the map to smaller resolution
and apply standard techniques to recover the power spectrum
at $l\leq 20-40$. The second limitation is common to all methods that compute
the power spectrum by means of the correlation function or in the presence of 
Galactic (and other) cut. The intrinsic multipoles 
need to be de-convolved from the tapering function, i.e., if
$C^{\rm intrinsic}_l$ are the multipoles of the sky radiation power 
spectrum, and $C_l$ are obtained using Eq.~(\ref{eq:cl}), then
\begin{equation}
C_l=\sum_{l^\prime}C^{\rm intrinsic}_{l^\prime}{\cal F}_{l-l^\prime}
\label{convolve}
\end{equation}
where ${\cal F}_l$ is the Legendre transform of the tapering function $F$. 

\begin{figure}[h]
\begin{center}
        \plotone{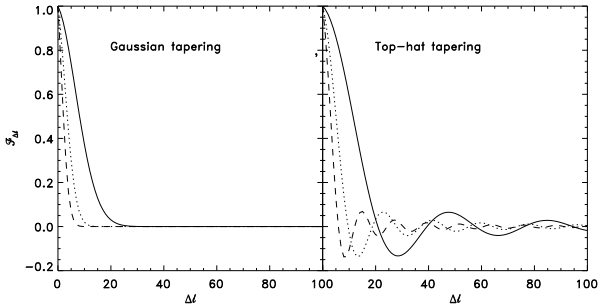}
\caption[]{Legendre transform of the tapering function (Eq. \ref{convolve}) 
plotted vs $\Delta l = l -l^\prime$. Left panel corresponds to Gaussian 
tapering, $F(\theta) \propto \exp(-\theta^2/\theta_t^2)$ and the right panel to 
the top hat tapering, $F(\theta) =1$ for $\theta \leq \theta_t$ and zero 
otherwise. Dashed, dotted and solid lines correspond to $\theta_t =18\deg, 
12\deg, 8\deg$ respectively.}
\label{figure3}
\end{center}
\end{figure}

Fig.~(\ref{figure3}) shows the Legendre transform of the tapering function as 
function of $\Delta l \equiv l - l^\prime$ for Gaussian and top-hat tapering. 
The advantage of Gaussian tapering is quite obvious as it leads to significantly 
less prominent side-lobes. Hence, we adopted it in 
our computations. As the figure shows, it would lead to FWHM correlation width of 
only $\Delta l \sim$ (a few) for tapering angles $\lsim 10\deg$. 
Correlations on the $C_l$'s would be dominated by tapering
and  correlated errors on $\xi_\nu$.

\section{\bf Application to WMAP}

WMAP is observing at 5 frequency bands: 23, 33, 41, 61,
and 94 GHz, (Jarosik et al. 2003, Bennett et al. 2003a).  
The beam response at
each band are given by a gain pattern, G, which is asymmetric and
non-gaussian. From the solid angle beam, one can always define a FWHM,
which, for increasing frequency channels, are equal to 
0.82, 0.62, 0.49, 0.33 \& 0.21 degrees, (Page et al. 2003).
 The sky maps based on the full two-years
of data are expected to have an $rms$ noise of $\simeq$35 $\mu$K per
$0.3\deg\!\times\!0.3\deg$ pixel.  By design, the noise will be
essentially uncorrelated from pixel to pixel and it is expected
to be highly Gaussian (Hinshaw 2000, Hinshaw et al. 2003).
  Due to the sky scanning strategy, the 
noise is reasonably uniform across the sky, except at the small regions near 
the ecliptic poles and at ecliptic latitude $\sim 45\deg$ where the
sensitivity will be somewhat higher. The WMAP radiometers produce raw 
temperature measurements that are the 
differences between two points on the sky separated $\sim 140\deg$ . Since a 
given pixel $i$ is observed with up to ~1000 different pixels $j$, the 
covariance between any given pair of pixels $(i,j)$ is much less than 1\% of 
the variance of pixel $i$.  The noise covariance in the final sky maps will, 
by design, be very nearly diagonal. 

\subsection{Homogeneous noise results}

\begin{figure}[h]
\begin{center}
        \plotone{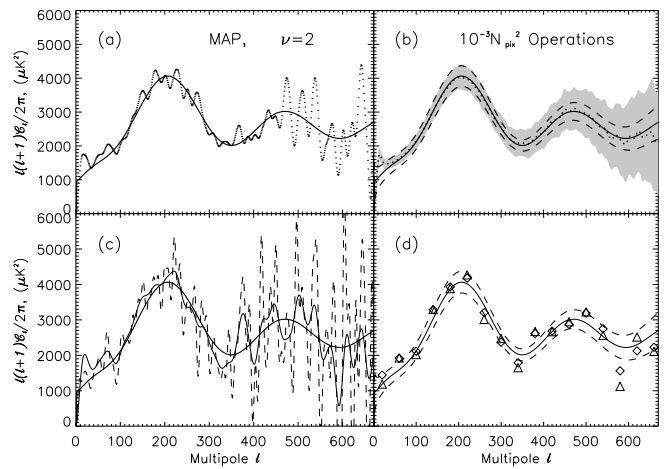}
\caption[]{
Radiation power spectrum for $\nu=2$ and different taper angles. 
In all panels the solid line shows the input power spectrum.
(a) Dotted line: power spectrum obtained from a single simulation 
with $\theta_t=12^0$. (b)
Dotted line: power spectrum  obtained from averaging 500 simulations,
shaded area: 1$\sigma$ error region obtained from the same simulations.
The dashed lines limit the 1$\sigma$ optimal variance error bars
(Eq.~(\ref{ov})). (c) The same as in (a) but for $\theta_t=8^0$
(solid line) and $\theta_t=18^0$ dashed line. (d) 
solid and  dashed lines as in (b); the symbols
displayed the band averaged power spectra with bandwidth $\Delta l = 40$
of the spectra plotted in (c): diamonds correspond to $\theta_t=8^0$
and triangles to $\theta_t=18^0$.
}
\label{figure4}
\end{center}
\end{figure}

First, we performed a set of 500 simulations for $\nu=(2,2.27,2.525)$
including only cosmological signal and white noise. 
For each simulation, we computed
the correlation function and the power spectrum.
Fig.~(\ref{figure4}) summarizes the results for the threshold $\nu = 2$. 
In panel (a) the dotted line shows the power spectrum recovered from a single simulation,
whereas the solid line gives the input power spectrum. Tapering angle was $\theta_t = 12\deg$.
In (b), we show the average results for 500 realizations:
the solid and dashed lines represent the input model and the average power spectrum
of the 500 simulations, respectively. The shaded area shows
the 1-$\sigma$ error bar region for each $C_l$ and the dashed lines
show the optimal variance. Our method traces
the acoustic peaks up to the beam scale, $l_{\rm beam} \sim 640$. Its accuracy
is roughly similar to the optimal variance $\sigma_{\rm OV}$ for $l\leq 300$. 
The spectrum of Fig.~(\ref{figure4}a) has been obtained in 
$10^{-3} N^2$ operations, 
using a taper window of $12\deg$ FWHM. 
The accuracy can be improved by lowering the threshold $\nu$ and hence
increasing the computational time. 

To see the effect of tapering, in Fig.~(\ref{figure4}c) we plot
the power spectra obtained for two different tapering angles for
the same simulation as in Fig.~(\ref{figure4}a). The solid
line corresponds to $\theta_t = 8\deg$ while the dashed line 
to $\theta_t = 18\deg$. Increasing the tapering angle leads to larger 
oscillations, but also decreases the correlations between different $l$.
In Fig.~(\ref{figure4}d) we plot the band power
average of the previous spectra: diamonds correspond to the tapering
angle of  $8\deg$ and triangles to $18\deg$. The solid line is the
input model and the dashed lines correspond to the optimal variance
error bars as in (b). As expected, both results are almost identical
since all the information at high $l$'s is contained at angular 
scales smaller than $1\deg$.

The correlation matrix, $ {\cal 
C}_{l,l^\prime}$, among different multipoles is given by:
\begin{equation}
{\cal C}_{l,l^\prime} =
\frac{ \langle \delta C_l \delta C_{l^\prime} \rangle}
{\sqrt{ \langle \delta C_l^2 \rangle \langle \delta C_{l^\prime}^2 \rangle}},
\label{eq:corcoeff}
\end{equation}
with $\delta C_l = C_l - \langle C_l \rangle$. The correlation coefficient 
matrix has been computed after using a Gaussian taper function $F(\theta) 
= \exp -(\theta / \theta_t )^2$. It is 
highly diagonal with FWHM width $\Delta l \sim 10$ for $\theta_t = 
8\deg$, {\em independently } of the value of the multipole $l$.
At $\Delta l\sim 20$ the correlation drops to the level of $ \sim 5\%
$,
independently of $\nu$.  Outside the central diagonal strip the residual 
correlations are due to shot noise from the finite number of simulations. 
For consistency, we repeated the analysis with only 125 simulations.
The width of the diagonal remained the same, but the off-diagonal terms
grew by a factor of two, consistent with Poisson statistics.

For larger/smaller tapering angles, the correlation scale $\Delta l$ will
be smaller/larger but, as demonstrated in Fig.~(\ref{figure4}d), it will not change
our estimate of the power spectrum. We also checked that our results did not
change if we use  the correlation function from the coarser (and computationally 
tractable) map from $\theta=10\deg$ out to $180\deg$ instead of tapering. 
The correlation on the final $C_l$'s are dominated by the correlated errors 
in $\xi_\nu$ at small angles.

\begin{figure}[h]
\begin{center}
        \plotone{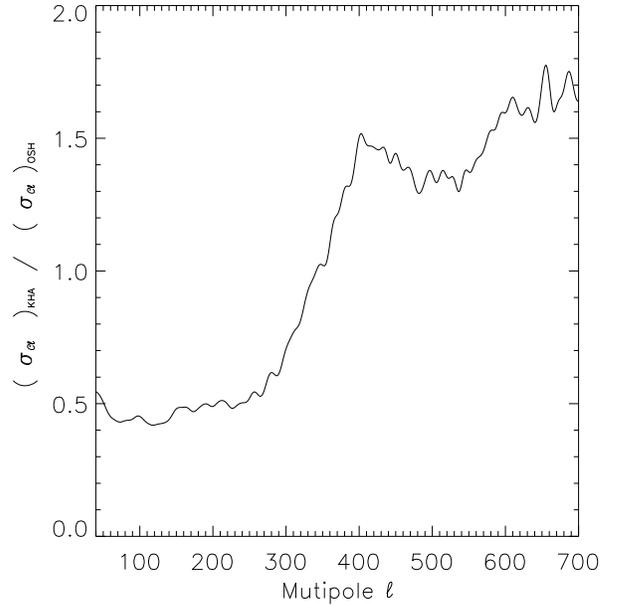}
\caption[]{Comparison of the error bars of OSH and KHA methods 
for a CMB map as seen by the
94 GHz channel of WMAP. We plot the ratio $ (\sigma_{C_l})_{\rm KHA} /
(\sigma_{C_l})_{\rm OSH}$. The uncertainties $\sigma_{C_l}$'s were computed
for raw multipole estimates $C_l$, without averaging. 
}
\label{figure5}
\end{center}
\end{figure}

In Fig.~(\ref{figure5}) 
we compare the accuracy of our method for WMAP and $\nu=2$ with
that of OSH. We compute the ratio of the uncertainties in the 
multipoles recovered by both methods.  Our method gives a comparable precision to that 
of OSH but requires much fewer operations. 
On the other hand, it gives fewer independent data points. In direct methods, such as OSH, 
the necessary Galactic cut leads to $C_l$'s that are
correlated on scales of $\Delta l \simeq 2-3$, while our method gives
a correlation length of about $\Delta l\sim  10$.

\begin{figure}[h]
\begin{center}
        \plotone{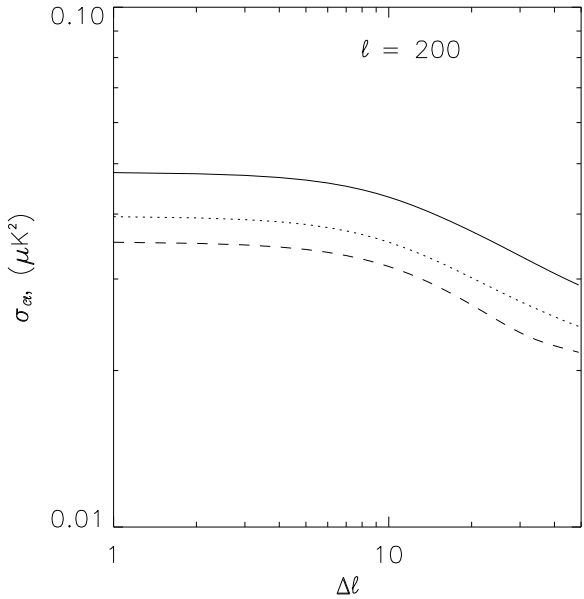}
\caption[]{
Scaling of band power error bar $\sigma_{C_l}$ with bandwidth $\Delta l$.
Solid, dashed and dotted lines correspond to $\nu=2.5,2.27$ and $2$, respectively.
}
\label{figure6}
\end{center}
\end{figure}

In Fig.~(\ref{figure6}) we show the variation of the error bar with 
bandwidth for the multipole at $l=200$. Solid, dashed and dotted lines correspond to
$\nu=2.5,2.27$ and $2$, respectively.
Due to the presence of this correlation among the multipole
estimates, this $\sigma_{C_l} \sim (\Delta l)^{-1/2}$ behavior is 
obtained for $\Delta l \gsim 10$, the correlation scale
introduced by our method. 

\begin{figure}[h]
\begin{center}
		\plotone{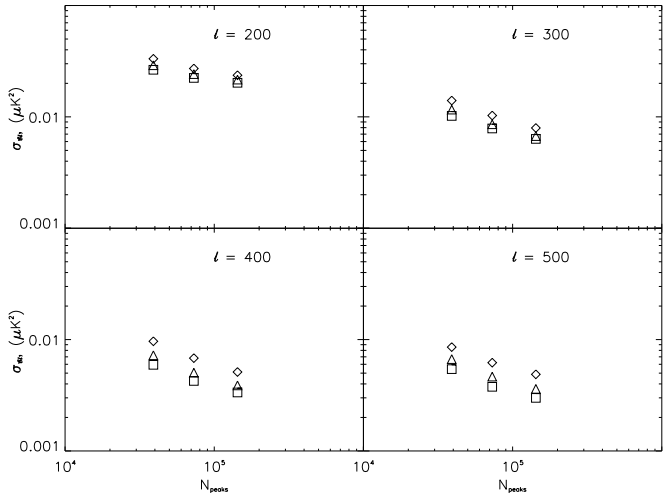}
\caption[]{ 
Band power error bars as a function of the
number of pixels used to compute the radiation power spectrum are shown
for four values of $l$. Diamonds, triangles and squares
correspond to  $\Delta l = 30, 50, 70$,  respectively. 
}
\label{figure7}
\end{center}
\end{figure}

In Fig.~(\ref{figure7}) we show $\sigma_{C_l}$ at $l=200, 300, 400, 500$
for the three thresholds. Smaller error bars correspond to 
smaller thresholds, i.e., to larger number of peaks. 
A power-law fit using all multipoles gives a power law behavior
of the form $\sigma_{C_l} \propto N_{\rm peaks}^{\beta}$, 
with $\beta = -0.405 \pm 0.016$. These relations  
can be used to estimate the amplitude of the error bar attached to
each multipole for a wide range of values of $\nu$ and $\Delta l$.
In particular, it can be used to find what values of $\nu$ and $\Delta l$
are necessary to achieve a given degree of accuracy in the power spectra
and the time required in the computation.

We
have shown that for a given experiment the accuracy of the $C_l$'s computed
spectrum depends mainly on one parameter: the peak threshold allowing to 
choose between larger accuracy or speed.

\subsection{Including foregrounds and realistic noise}

We now generalize the method to include realistic foreground 
emission and inhomogeneous noise and will demonstrate that also in this case
the KHA method gives similar estimates of the CMB power spectrum. 

\subsubsection{Foregrounds}

After the first year data release, the WMAP team made public foreground maps
built after combining the five different frequency maps provided by the 
instrument, (Bennett et al. 2003b). These foreground maps were constructed
after  applying a Maximum Entropy Method using existing templates as priors, 
and the resulting model for galactic emission matches the observed emission
to $<1$\%. Using this model for Galactic emission, we get that with the
 fiducial value of 20\% for the amplitude of the foreground residuals above 
$b\! = \!10\deg$, they are much smaller
than the intrinsic CMB temperature fluctuations associated with peaks 
($\delta T \gsim 200 {\rm \mu}$K for $\nu=2$).

To model the effect of foregrounds, we used simulated maps provided to us earlier by 
G. Hinshaw of the WMAP science team (2002, private communication).
These were produced by combining the
Haslam 408 MHz map and the Schlegel, Finkbeiner \& Davis (1988, hereafter SFD) 
IRAS/DIRBE 100 ${\rm \mu}$m dust 
map. The Haslam map was used as the template for synchrotron emission and
the SFD map as the template for dust and free-free.
These maps were scaled to microwave frequencies using the COBE DMR-based fits of
these templates (with 7 degree resolution).  The frequency by frequency fit 
results are in Table 1 of Kogut et al. (1996). In detail, the 
Haslam map was scaled using a power law index $\alpha_{\rm syn} = -3$.
The SFD free-free map was scaled using $\alpha_{\rm ff} = -2.15$ and
the SFD dust model was scaled using an index $\alpha_{\rm dust}= 2.0$.
This model is known to over-predict the plane emission at DMR resolution,
thus it is likely to be conservative.
In the foreground maps, we did not include point sources. 
Sokasian, Gawiser \& Smoot (2001) have compiled and analyzed the available
extra-galactic point source data and have concluded that such sources will
contribute negligibly to the angular power spectrum at 94 GHz for $l\! < \!800$.

\begin{figure}[h]
\begin{center}
        \plotone{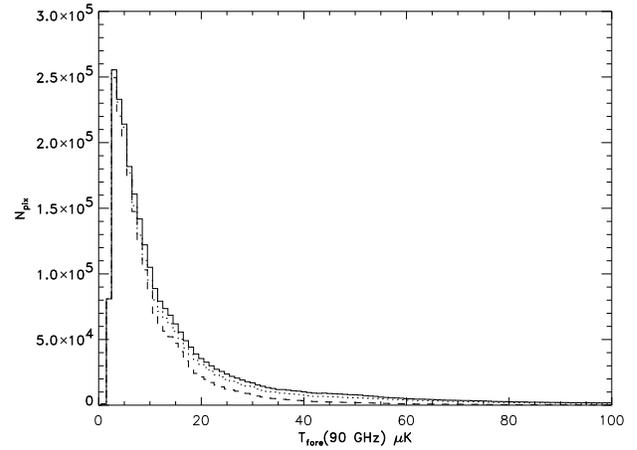}
\caption[]{The histograms of the 94 GHz foreground emission with WMAP resolution. 
Solid, dotted and dashed lines  correspond to  
Galactic cut $b_{\rm cut}=5,10,20$ degrees, respectively.
}
\label{figure8}
\end{center}
\end{figure}

Fig.~(\ref{figure8}) shows the histogram of foreground contributions to the WMAP 
94GHz channel for 3 values of Galactic cut: the solid, dotted
and dashed  lines correspond to $|b|_{\rm cut} = 
5\deg$, $10\deg$ and $20\deg$, respectively. For 
comparison at $\nu=2$ the value CMB contribution to the remaining pixels will be 
$\sim 200 {\rm \mu}$K. Clearly, foregrounds are not likely to significantly affect the method.

\begin{figure}[h] 
\begin{center} 
        \plotone{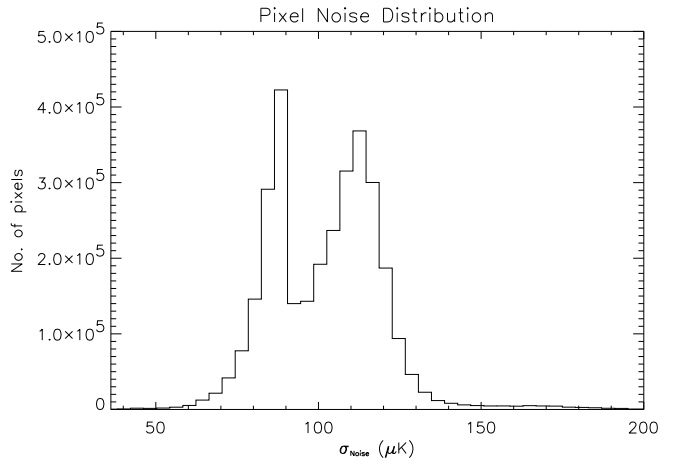} 
\caption[]{Histograms of pixel noise variance for the WMAP 94 GHz channel 
with pixels size of $7$ arcmins.
} 
\label{figure9} 
\end{center} 
\end{figure}

\subsubsection{Noise model}

Our statistic $\xi_{\nu}$ has been worked out for
homogeneous Gaussian fields. The inversion of $\xi_{\nu}(C(\theta )/C_0)$
into the correlation function $C(\theta )$ is accurate provided that
signal and noise correlation functions are uniform across
the sky. For most experiments this is not the case as the sky coverage
is not homogeneous and the noise variance changes with location.
For this more realistic case we adopt the following strategy:
if the noise variance $\langle {\cal N}_i {\cal N}_i \rangle$ changes
according to the number of times a pixel has been observed
($\langle {\cal N}_i {\cal N}_i \rangle\sim t^{-1/2}_{\rm observ}$), pixel $i$ will be
selected as peak above a threshold $\nu$ if it verifies 
$|\delta T_i | \ge \nu  (C^{\rm CMB}(0) + \langle {\cal N}_i {\cal N}_i \rangle)^{1/2}$,
i.e., we take into account the local noise variation.
In the expression, $C^{\rm CMB}(0)$ is the variance of the cosmological signal.
We apply this formalism to the noise model of the WMAP two-year scanning 
strategy. This model gives the number of 
observations ($N_{\rm obs}$) in each pixel at the end of the second mission
year. Fig.~(\ref{figure9}) shows the 
distribution of the r.m.s. noise distribution for WMAP 94 GHz channel, for 
$ 7$ arcmin-sized pixels. In the figure, the noise average is 
$\sqrt{\langle {\cal N}_i {\cal  N}_i   \rangle_{\rm full\; sky}} \simeq 97\mu$K per pixel.

\begin{figure}[h]
\begin{center}
        \plotone{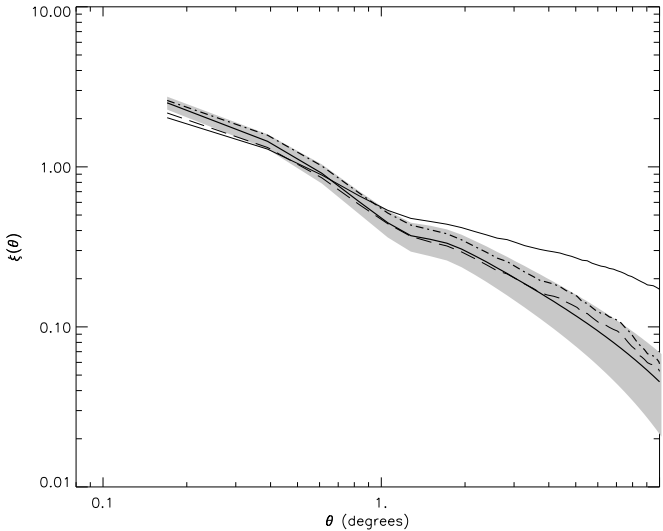}
\caption[]{
Thick solid and dot-dashed lines: $\xi_{\nu=2}$ of the input model and estimator
when the noise is homogeneous, as in Fig.~(\ref{figure1}).
When the noise is inhomogeneous, the thin solid line gives $\xi_{\nu=2}$
obtained when we apply the same method as if the noise were homogeneous.
The long-dashed line corresponds to the case when we take into account 
the variable amplitude of the noise and select pixels accordingly (see text).
The shaded area is the 1-sigma optimal variance error bars.
}
\label{figure10}
\end{center}
\end{figure}

Fig.~(\ref{figure10}) shows that this gives similar results
as for homogeneous noise.
The thick solid line corresponds to the theoretical estimate
of $\xi_{\nu=2}$ according to eq~(\ref{xi}). For a fixed cosmological
signal, if we add homogeneous noise to a CMB map, the dot-dashed line will
be the numerical estimate, like in Fig.~(\ref{figure1}). The thin solid line
gives the  estimate of $\xi_{\nu=2}$ assuming the noise
is homogeneous when it is not, while the long-dashed line gives the estimate
when the pixels have been selected according to the criterion defined
in the previous paragraph.  The latter result is comparable to the case
of homogeneous noise and, in this sense, this procedure gives
an optimal estimator.

\subsubsection{Results}

We performed 150 simulations of WMAP data using the SCDM model,
with inhomogeneous noise and foreground residuals.
We choose three different galactic cuts and two different amplitudes
for the foreground residuals: 10\% and 20\% amplitude of  the original
foreground data at 94GHz.  In the first case, we performed simulations
with a Galactic cut at $|b|=5\deg$ and $10\deg$ and for the 20\% residuals 
we imposed a cut at $|b|=20\deg$. We have checked that these models 
give a foreground residual level comparable to the actual
contribution found in the data by Bennett et al. (2003c).

\begin{figure}[h]
\begin{center}
        \plotone{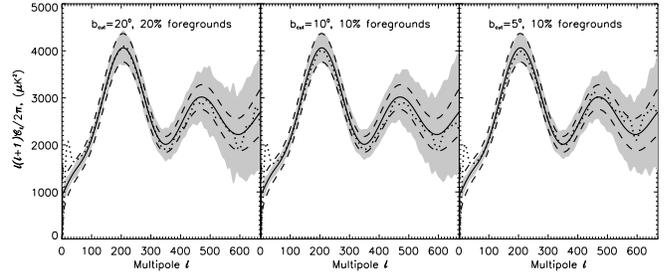}
\caption[]{
Radiation power spectra obtained from maps containing inhomogeneous
noise and foreground residuals. The correlation function
was computed using all pixels above $\nu = 2$.
As in Fig.~(\ref{figure4}), solid lines correspond to the input
power spectrum. The dotted lines  give the average of 150 simulations.
The dashed lines show the 1-$\sigma$ confidence
level given by the optimal variance of Eq.~(\ref{ov}). 
The shaded area corresponds to the 1-$\sigma$ error region obtained from the
simulations. 
}
\label{figure11}
\end{center}
\end{figure}

In Fig.~(\ref{figure11}) we show the results of 150 simulations.
As in Fig.~(\ref{figure4}), the dotted lines shows the mean power 
spectrum of all the simulations, while the solid line shows the input 
model. Shaded areas give the 1-$\sigma$ confidence level and
dashed lines the optimal variance 1-$\sigma$ error bars (see Eq.~(\ref{ov})).
In the whole range, the error bars were almost identical (less than 5\% increase)
to the case with no foregrounds and homogeneous noise.
The correlations in $l$-space are practically the same as before ($\Delta l\simeq 10$).
The deviations of the mean from the input
model are from the smaller number of simulations.

To quantify the amplitude of the non-Gaussian components, we
computed the skewness and kurtosis of our simulations. The skewness and 
kurtosis are defined (cf. Press et al. 1992) as $\sigma_3=\langle (\delta T/
\sigma_2)^3 \rangle$, $\sigma_4 = \langle (\delta T/\sigma_2)^4 \rangle - 3$,
respectively, with $\sigma_2$ being the variance of the data. 
In our case, the non-gaussian signal is dominated by the foreground
residuals. We computed the skewness and kurtosis for a randomly selected map
 of the previous set of simulations. 
For a Galactic cut at $10\deg$ and 10\% foreground 
residuals amplitude, the values found were $\sigma_3 = 0.12$ and 
$\sigma_4 = 1.2$.  For comparison, for the same CMB map re-simulated
with a realistic noise component, galactic cut and no foregrounds, 
the values were $\sigma_3 = 0.02$ and $\sigma_4 = -0.03$.

To conclude, the addition of a non-Gaussian signal
(the foreground residuals) and inhomogeneous noise does not 
affect significantly the performance of the method.
The reason that the introduction of Galactic foregrounds produces
only small variations in the results is at the very core
of our procedure: with the new KHA statistic, Eq. \ref{xi}
we select regions with high 
S/N ratio, the peaks, and small departures from gaussianity neither 
degrades the quality of our subset of data, nor introduces additional
correlations.  

The results presented in Fig.~(\ref{figure11}) would improve for lower thresholds. 
As the method provides correlated $C_l$'s, it is necessary to bin the estimates into band
powers. Other methods share, to some extent, this limitation; 
direct computation of the power spectrum 
from a map which necessarily has a cut due to bright foreground regions, is 
possible only with a finite band-width determined by size of the Galactic 
cut. This was avoided in the COBE/DMR data with the Gramm-Schmidt 
orthogonalization of the base functions (G\'orski 1994), but the procedure 
is impractical for the mega-pixel CMB datasets. For the WMAP 94 GHz channel, 
our method
provides, under reasonable accuracy, around 12-14 independent data points 
for the first two Doppler peaks of the CMB power spectrum, and with a gain 
factor of  $10^3-10^4$ in CPU time compared to standard methods.

We have also demonstrated that our method retains its accuracy in the 
presence of foregrounds 
and for the realistic/inhomogeneous WMAP noise.  The treatment in the presence 
of the inhomogeneous noise can also be improved by removing the pixels with 
significantly fewer observations. Furthermore, because the peaks  
method computes a correlation function, $\xi_\nu$, it is immune to masking.
The latter would allow us to reduce the foregrounds signal by removing from
 the CMB maps more isolated regions with higher foreground contribution. 

\section{\bf Application to Planck} 

The Planck satellite\footnote{(http://astro.estec.esa.nl/Planck/)}
is due to be launched  
in early 2007 to map the all sky CMB distribution with an even finer 
resolution than WMAP. 
It will have two instruments: the low-frequency instrument (LFI), that  
operates at three frequency channels between 30 and 70 GHz, and the  
high-frequency instrument (HFI) in six frequency channels between 100 and 857  
GHz. The highest Planck resolution will be 5 arcmin and the instrument noise  
after one year of operations is expected to be around 6-10 $\mu$K. 

\begin{figure}
\begin{center}
        \plotone{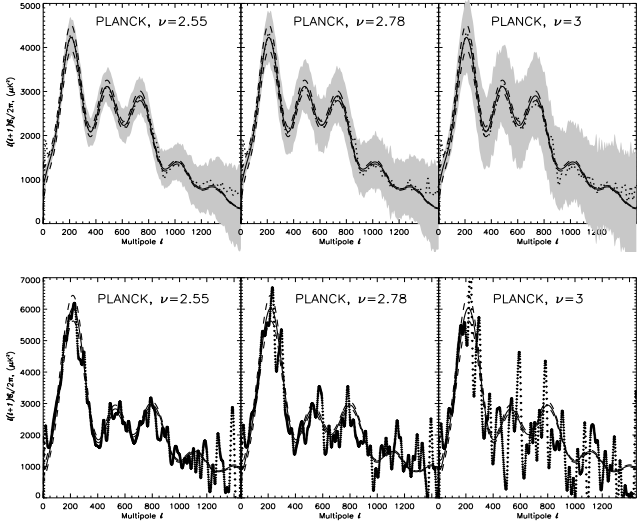}
\caption[]{Radiation power spectrum for the Planck 217 GHz channel: in the
top row we have considered the SCDM model used in WMAP simulations.
In the top panels we follow the conventions of Fig.~(\ref{figure4}):
solid lines represent the input model, dotted lines the mean
of 125 simulations, dashed lines are optimal variance error
bars at $1\sigma$ and the shaded areas are the same confidence limit
obtained from the simulations. In the bottom row, we show results for the
concordance $\Lambda$CDM model for {\em one} single realization 
(dotted line). Again, the input
model is given by the solid line and the optimal variance error bars are
given, as before, by the dashed lines.
} 
\label{figure12}
\end{center}
\end{figure}

For this case 
we simulated the 217 GHz channel, with a beam of $5.5'$ FWHM and 
an {\em average} noise level of $11.4{\rm \mu}$K per beam area.
We tested this channel in two different cosmological models: for consistency,
we performed 125 simulations with the same SCDM model used for WMAP, (top
row of Fig.~(\ref{figure12}), but we also tested our method with
the $\Lambda$CDM concordance model, given by $\Omega_{\rm m}=0.292$,
$\Omega_{\Lambda}=0.708$, $\Omega_{\rm b}=0.044$ and $h=0.72$,
(bottom row of the same figure).
We assumed that the instrumental noise was the sum of
two different components with  white noise and 
$1/f$ contributions
(see Maino et al. 1999 for a detailed account of systematic effect
on the Planck LFI instrument).

The $1/f$ component is characterized by a {\it knee frequency $f_k$},
for which the power spectra of both white and $1/f$ noise components are equal.
If the spin frequency $f_s$ is not much smaller than the knee frequency $f_k$,
then we can use the fact that the telescope spends sixty minutes in each ring
to simply consider the average noise pattern for every ring, (Maino et al.
2002). In other words, we can
neglect the noise high frequency components present in the same ring, as these
will be averaged out. 
We model the $1/f$ component as a different baseline
present in every scanned ring, giving a pattern of stripes in the
noise map (Janssen et al. 1996). We used the Planck 217 GHz 
channel noise
model, provided to us by the Max Planck Institute f\"ur Astrophysik (MPA)
at Garching,
that includes all systematic effects. 
For each simulation, we added a realization of this $1/f$ component to the
 cosmological and white noise components. For simplicity we did not include
 foreground residuals as in the previous section we have shown to give 
negligible contribution. For similar reasons we also did not include point 
sources.

The maximum multipole to which a beam of $5.5'$ FWHM 
is sensitive is $l_{\rm beam}=1472$. The HEALPix 
configuration chosen to pixelize the CMB sky as seen by this channel was 
$N_{\rm side}=1024$, which yields pixels of 3.5 arcmins. 
In Fig.~(\ref{figure12}) we show the results 
of 500 simulations. As indicated, we have chosen three threshold levels  
that have the same number of pixels that the three $\nu$'s considered for WMAP.
We plot the results for those
thresholds following the conventions of Fig.~(\ref{figure4}).  
Note that, while we obtain the radiation power spectrum 
from the same number of pixels as in the case of WMAP, the method
recovers it up to $l=1472$, the largest possible multipole 
determined by the beam size.
This could be expected, since the clustering of peaks and troughs
is a statistic particularly sensitive to small angular scales. In the range
$\theta \in [\theta_{\rm beam},\sim 10^0]$, or equivalently 
$l\in[40,l_{\rm beam}]$,
our method reconstructs the power spectrum with an accuracy
depending only on the threshold. The range in $l$-space probed by the
method depends on the resolution of each experiment but is
{\it independent} of the total number of peaks used in the analysis.
 
Using the data from the three set of simulations we can compute
how the band power error bars scale with the threshold $\nu$.
In this case, a fit of the form $\sigma_{C_l}\propto N_{\rm peaks}^\beta$ 
to the error bars of all multipoles gives: $\beta= -0.65\pm 0.011$, close to 
the expected behavior $\sigma_{C_l}\propto 1/\sqrt{N_{\rm peaks}}$. 
As mentioned before,
this scaling can be used to decide what threshold to choose to
achieve a preselected accuracy.  The correlations on the $C_l$'s
were similar to those of WMAP, of the order of $\Delta l \simeq 10$.

\section{Discussion and conclusions}

We have presented the detailed implementation of the peaks KHA method for 
computing the CMB power spectrum. The method is based on the
correlation properties of peaks on the CMB and assumes that temperature 
fluctuations are Gaussian. We have shown, that the method is robust
in the presence of Galactic (non-Gaussian) foregrounds and realistic
WMAP and Planck noise inhomogeneities.

\begin{figure}
\begin{center}
        \plotone{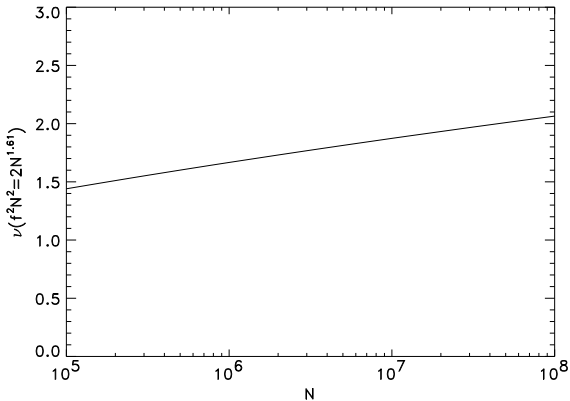}
\caption[]{
The value of the peak threshold where the number of operations in the KHA 
method equals $N^{1.61}$, is plotted as a function of the number of pixels.
} 
\label{figure13}
\end{center}
\end{figure}

This method provides a new and independent way to
characterize the sub-degree structure of the CMB. Its accuracy is generally lower 
than standard methods, but it is remarkably faster for the range of
values of $\nu$ required by WMAP and Planck. This is demonstrated in 
Fig.~(\ref{figure13}), where we
compare speed of a standard ${\cal O}$(N log N) $\sim ~N^{1.61}$ method
with ours.
 
For WMAP the KHA method should work most 
optimally for peak thresholds of $\nu\sim 1.8-2.5$ or in only $(2.5\times 
10^{-3} - 8\times 10^{-5})N^2$ operations.
Applying the method to the simulated PLANCK sky maps has shown that we could 
probe, with the same number of points,
the power spectrum out to much higher multipoles and without 
significant loss of accuracy. For 
Planck mission parameters, the KHA method should work most optimally for  
$\nu\sim 2-3$, reducing the number of steps down to $\sim 
4\times 10^{-6}N^2$ operations for thresholds as high as $\nu \sim 3$. 
This can potentially lead to a gain of up to $\sim 10^5$ in speed compared to 
the existing O$(N^2)$ methods.

A significant advantage is that because we use a 
$\xi_{\nu}$ statistic, the computed 
CMB spectrum is immune to masking, allowing to remove high noise and 
foreground emission parts of the sky. This also makes straightforward the 
analysis of particular isolated regions in the sky, enabling its 
application to future CMB experiments of very high angular resolution 
scanning small celestial patches.

\begin{figure} 
\begin{center} 
        \plotone{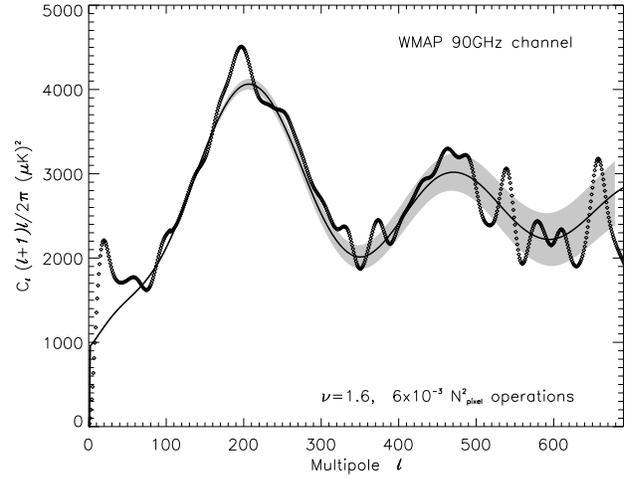} 
\caption[]{
Radiation power spectrum that would be recovered using the WMAP second
year data with a threshold of $\nu=1.6$: it was obtained in
only $6\times 10^{-3}N^2$ operations.
The solid line corresponds to the input model and the dotted line
is the raw power spectrum recovered from one single simulation, prior to
binning.
The shaded area shows the extrapolated 1-$\sigma$ uncertainties of
the CMB power spectrum binned with $\Delta l=45$. 
}
\label{figure14} 
\end{center} 
\end{figure}

With the scalings of $\sigma_{C_l}$ with $\nu$ for both WMAP and
Planck one can estimate the uncertainties of the different multipoles at 
still lower peak threshold values. 
In Fig.~(\ref{figure14}) we show what would be the spectrum
recovered from the 2 year WMAP data release using
a threshold $\nu=1.6$.  The whole calculation would require only 
$6\times 10^{-3}N^2$ operations.
 
\begin{acknowledgements} We are grateful to Gary Hinshaw for 
providing us with the Galactic foregrounds 
model data and the WMAP observations model used in the simulations. We also
thank the Planck simulation center at the Max Planck Institute f\"ur
Astrophysik for allowing us to use their 217 GHz noise model for Planck. 
Some of the results of this paper have been derived using the HEALPix
package, (G\'orski, Hivon \& Wandelt 1999).
C.H.M. and F.A.B. acknowledge support of Junta de Castilla y Le\'on (project
SA002/03) and Ministerio de Educaci\'on y Cultura (project BFM2000-1322).
C.H.M acknowledges the financial support provided through the European
Community's Human Potential Programme under contract HPRN-CT-2002-00124, 
CMBNET. C.H.M. also thanks the Astrophysikalisches Institut Potsdam for 
allowing the use of their computer resources and J.A.Rubi\~no-Mart\' \i n for 
encouraging the final submission of the manuscript.

\end{acknowledgements}

{\bf REFERENCES}\\
Bardeen, J.M., Bond, J.R., Kaiser, N. \& Szalay, A.S. 1986, ApJ, 304, 15\\
Bennett, C., Halpern, M., Hinshaw, G. et al. 2003a, ApJ, accepted\\
Bennett, C., Hill, R.S., Hinshaw, G. et al. 2003b, ApJ, accepted\\
Bond, J.R., Jaffe, A.H. \& Knox, L. 2000, ApJ, 533, 19.\\
Finkbeiner, D.P., Davis, M., Schlegel 1999, ApJ, 524, 867\\
G\'orski, K.M. 1994, ApJ 430, L85\\
G\'orski, K.M, Hivon, E. \& Wandelt, B.D., 1999, Proceedings of the MPA/ESO
Cosmology Conference "Evolution of the Large Scale Structure",
eds. A.J.Banday, R.S.Sheth and L.Da Costa, PrintPartners Ipskamp, NL,
pp.37-42.\\
Hinshaw, G. 2000, Proceedings of the MPA/ESO/MPE Workshop ``Mining the Sky'',
held at Garching, Germany, 31st July - 4th August 2000. Edited by A.J.Banday,
S.Zaroubi and M.Bartelmann. Heidelberg: Springer-Verlag, 2001, p.447 \\
Hivon, E., G\'orski, K.M, Netterfield, C.B. et al. 2002 ApJ, 567, 2\\
Hu, W. \& Dodelson, S. 2002, A.R.A.A., 40, 171\\
Janssen, M., Scott, D., White, M. et al. 1996, astro-ph/9602009\\
Jarosik, N., Bennett, C.L., Halpern, M. et al., 2003, ApJS, 145, 413\\
Jensen, L.G. \& Szalay, A.S. 1986 ApJ, 305, L5\\
Kaiser, N. 1984 ApJ, 282, L9\\
Kashlinsky, A. 1987, ApJ 317, 19\\
Kashlinsky, A., Hern\'andez-Monteagudo, C. \& Atrio-Barandela, F. 2001
ApJ, 557, L1 (KHA)\\
Knox, L., 1995, Phys. Rev. D, 52, 4307\\
Kogut, A., Banday, A.J., Bennett, C.L. et al. 1996, ApJ, 464, L5\\
Landy, S.D \& Szalay, A.D. 1993 ApJ 412, 64\\
Maino, D., Burinaga, C., Maltoni, M. et al. 1999, A\&AS, 140, 383\\
Maino, D., Burinaga, C., G\'orski, K.M, Mandolesi, N. \& Bersanelli, M.
 2002, A\&A, 387, 356\\
Oh, S.P., Spergel, D.N. \& Hinshaw, G. 1999 ApJ, 510, 551\\
Padmanabhan, N., Seljak, U. \& Penn, U.L. 2003 NewA 8 581\\
Page, L., Nolta, N.R, Barnes, C. et al., 2003, ApJ, accepted\\
Peebles, P.J.E. 1980 ``The Large Scale Structure
of the Universe'', Princeton University Press, Princeton\\
Penn, U., 2003, astro-ph/0304513\\
Press, W.H., Teukolsky, S.A., Vetterling, W.T. \& Flannery, B.P. 1992 
``Numerical Recipes'', 2nd edition, Cambridge University Press, Cambridge\\
Rice, S.O. 1954, in ``Noise and Stochastic Processes", ed. Wax, N., p.133 
Dover (NY)\\
Seljak, U. \& Zaldarriaga, M. 1996, Ap. J., 469, 437
Smoot, G.F., Bennett, C.L., Kogut, A. et al. 1992, ApJ, 396, L1\\
Sokasian, A., Gawiser, E. \& Smoot G.F. 2001, ApJ 562 885\\
Szapudi, I., Prunet, S., Pogosyan, D.  et al 2001, ApJ, 548, L115\\
Szapudi, I., Prunet, S. \& Colombi, S. 2001, ApJ, 561, L11\\
Schlegel, D.J., Finkbeiner, D.P. \& Davis, M. 1998, ApJ, 500, 525\\
Tegmark, M. 1997, Phys.Rev.D., 55, 5898\\
Wandelt, B.D., Hivon, E. \& G\'orski, K. 2001, Phys.Rev.D, 64, 083003 \\

\end{document}